\documentclass[preprint,showpacs,preprintnumbers,amsmath,amssymb]{revtex4}

\usepackage{graphicx}
\usepackage{dcolumn}
\usepackage{bm}

\begin{document}

\title{Magnetization dynamics with a spin-transfer torque \\
}

\author{Z. Li and S. Zhang}
\affiliation{
Department of Physics and Astronomy, University of
Missouri-Columbia, Columbia, MO 65211
}

\date{\today}

\begin{abstract}  
The magnetization reversal and dynamics of a spin valve pillar, whose
lateral size is 64$\times$64 nm$^2$, are studied by using micromagnetic 
simulation in the presence of spin transfer torque. Spin torques display both
characteristics of 
magnetic damping (or anti-damping) and of an effective magnetic field.
For a steady-state current , 
both M-I and M-H hysteresis loops show unique features, including 
multiple jumps, unusual plateaus and precessional states. 
These states originate from the  
competition between the energy dissipation due to Gilbert damping and 
the energy accumulation due to the spin torque 
supplied by the spin current. The magnetic energy 
oscillates as a function of time even for a steady-state 
current. For a pulsed current, the minimum 
width and amplitude of the spin torque for achieving
current-driven magnetization reversal are quantitatively determined. 
The spin torque also shows very interesting thermal activation that is
fundamentally different from an ordinary damping effect.
\end{abstract}

\pacs{  }

\maketitle

\section{Introduction}

Recently, there is considerable interest in the phenomenon of 
spin-polarized current induced magnetization switching. This 
phenomenon was first suggested by Berger \cite{Berger} and 
Slonczewski \cite{Slonczewski}, based on a rather
general argument: for a system consisting of itinerant electrons and 
local moments, the total angular momentum 
is conserved even for the system out of equilibrium, and thus the
divergence of the spin current of itinerant electrons must
be accompanied by an equal and opposite change of the angular momentum of
the local moment. This change is equivalent to a spin torque
acting on the local moment. Since then,
there are many theoretical \cite{Bazaliy, Waintal, Stiles, 
Brataas, Zhang, Slonczewski2} 
and experimental efforts \cite{Tsoi, Myers, Sun1,
Katine, Grollier, Albert} to understand the 
microscopic origins. In this paper, we do not discuss the microscopic
origins and the detailed formalism of the current-induced spin torque. 
Instead, we consider the consequences of the spin torque on the 
dynamics of the local moment of a spin valve structure.

Regardless of the detailed physics involved microscopically, the effect of
the spin angular momentum transfer can be captured by an additional
term in the macroscopic Landau-Lifshitz-Gilbert (LLG) equation,
\begin{equation} \label{LLG} \frac{d{\bf M}}{dt}= - \gamma {\bf M}
\times {\bf H}_{eff}+ \frac{\gamma a_J}{M_s} {\bf M}\times ({\bf M}
\times \hat{\bf M}_p )+ \frac{\alpha}{M_s} {\bf M}\times \frac{d {\bf M}}{dt},
\end{equation}
where $\gamma$ is the gyromagnetic ratio, ${\bf M}$ is the magnetization
vector of the free layer, $\hat{\bf M}_p$ is the {\em unit} vector whose
direction is along the magnetization of the pinned layer, 
$M_s $ is the saturation magnetization,
${\bf H}_{eff}$ is the effective
magnetic field including the external field, the anisotropy field,
the exchange field, the demagnetization field and the random thermal field
whose form will be discussed in the next section. The term proportional
to $a_J $ is the novel spin torque term, and the last term is the 
Gilbert damping term. It is noted that $a_J$ has dimensions of magnetic field.

Experimental verification of the spin torque has been carried out
in magnetic nanowires \cite{Wegrowe}, spin valve pillar 
structures \cite{Myers, Katine, Grollier}, point-contact 
geometry \cite{Tsoi, Tsoi2}, 
and magnetic tunnel junctions \cite{Sun1, Sun3}.
The convincing observation of these experiments is that there exists a
critical current density above which the magnetization can be switched back and
forth. Other properties, such as thermal effects \cite{Myers2}, 
also agree with the spin torque term in Eq.~(1). Theoretical analysis   
has been mostly confined to a single domain
structure \cite{Sun2,Bazaliy2}. Both analytical and numerical solutions
indicate that the magnetization reversal becomes very complicated even
for a single domain structure. For example, the hysteresis for a fixed current
density may display a precessional mode, and there is a possibility of 
an inverse hysteresis (the magnetization along the direction of applied field
decreases when one increases the applied field). 
Miltat et al. \cite{Miltat} made a step forward in relaxing the single
domain assumption by studying the magnetization reversal for an ``S'' state 
and a ``leaf'' state \cite{Cowburn}; these are typical equilibrium 
domain structures of a submicron thin film. 
They concluded that the effect of the current is quite
different for these two structures even though they have nearly the same
magnetic energy.

The spin torque is fundamental new from the pure precessional term
(first term in Eq.~(1)) and from the damping term (last term in Eq.~(1)).
The precession term conserves the magnetic energy and it determines
the precessional frequency of the magnetization dynamics. The damping
term makes the magnetic system relax to a local energy minimum, i.e.,
it dissipates the energy during magnetization dynamics. The spin torque
term, however, can have both effects: it can be a source of precessional
motion as an effective field and it can serve as damping (or anti-damping)
sources. It is this dual function that motivates us to 
study magnetization dynamics in a realistic magnetic
nanostructure. In this paper, we perform an extensive study on 
the effect of the spin
torque from the LLG equation, Eq.~(1). The paper is organized 
as follows. In Sec.~II, the model geometry of 
a spin valve pillar is defined and the method for micromagnetic simulation 
is outlined. Next, we calculate hysteretic and dynamical properties 
in the presence of the spin torque; in particular, we analyze precessional
states in detail. We also present the results for a 
non-steady-state current in Sec.~III. Finally we summarize what
are the most interesting features of magnetization dynamics 
in the presence of the current-induced spin torque. 

\section{Micromagnetic model with the spin torque}

To mimic the experimental geometry performed by the group at 
Cornell \cite{Katine}, we define
our structure in Fig.1. The electrical current flows perpendicular to the 
plane of the layer. The coordinate axes are so 
chosen that the $x$ and $y$ axes are in the plane of the layer and the 
$z$-axis is perpendicular to it. A spin-polarized current 
$j$ enters the spin valve pillar in the $z$ direction and 
we assign the positive value of the parameter 
$a_J$ in Eq.~(1) for the current flowing from the thicker ferromagnetic 
layer to the thinner one. The current induced magnetic field 
is ignored.

Since we focus our study on magnetization dynamics of the free layer, 
we assume that the pinned ferromagnetic layer is held fixed at the 
direction of the easy axis of the free layer, i.e., 
in positive $x$ direction.
The following materials parameters for the free layer are used:
the lateral size is 64nm$\times$64nm, the thickness is 2.5nm, the 
uniaxial anisotropy field $H_K$ is $500$(Oe), and the saturation magnetization
$4 \pi M_s = 12,000$ (Oe). These parameters are reasonably consistent with
the experiments by the Cornell group \cite{Katine}. We note that the choice
of the small layer thickness has two advantages: the magnetization
direction in the thickness direction will be uniform, i.e., a 2-d 
micromagnetics modeling is sufficiently accurate and the spin torque per 
unit volume is large for a fixed current density (since the total spin 
torque will be almost independent of the free layer thickness 
within the applicability of several theoretical models). 

With above specified parameters, we begin our simulation by laterally 
dividing the free layer into an N$\times$N grid. In most cases presented in
the paper, we choose N=16 so that the grid spacing is $D=4.0nm$.
Our goal is to calculate magnetization dynamics as the external magnetic
field, the spin torque, and the damping parameter vary. The magnitude
of the spin torque is proportional to the current density. Here we have chosen 
the unit of $a_J$ in Oesteds. The numerical value of $a_J$ 
has been estimated \cite{Slonczewski,Zhang}:  
for the current density $j=10^8 A/cm^2$,
$a_J $ in Eq.~(1) is about 1 kOe. This conversion between $a_J$ and the
current density is, however, irrelevant for us since we have written our 
results in terms of $a_J $ and thus we do not specify absolute values
of the current density. The Gilbert damping constant $\alpha$ in Eq.(1)
is not accurately known in a spin valve system. Among many
physical sources for the damping, there is an additional contribution
at the interface. Recent studies \cite{Back,Bret,Tserkovnyak} 
for ultrathin films show that the damping constant
$\alpha$ is much enhanced when a nonmagnetic metal is deposited on the
ferromagnetic film. For $Cu/Co$ bilayers, the damping constant is considerably 
larger than that of bulk value of $Co$. With this difficulty in choosing 
$\alpha$, we will vary it from 0.001 to 0.2 to address the 
damping dependence of the magnetization dynamics. 

The effect of temperature on the dynamical behavior is also 
included in our simulation by adding a random thermal 
field to the effective magnetic field. The thermal field 
$H_{th,i}^{\xi} (t)$ at each site $i$ is assumed to be
an independent Gaussian random function with its zero mean and 
no correlation, i.e., $<H_{th,i}^{\xi} (t)H_{th,j}^{\eta} (t')>=
2G\delta_{ij} \delta_{\xi \eta} \delta (t-t')$, where $i, j$ are the cell 
indexes, $\xi$, $\eta$ represents three Cartesian components, 
$G=\alpha k_{B}T / \mu_{0} \gamma M_{s} \nu$, and  
$\nu$ denotes the discretization volume of the computational cells.
By adding the thermal activation, the 
Landau-Lifshitz-Gilbert 
equation is converted into a stochastic differential 
equation with multiplicative noise. The integration of the 
stochastic Landau-Lifshitz-Gilbert equation 
is performed by employing the stochastic Heun method, 
by starting from a given initial configuration, and 
updating recursively the state of the system, 
${\bf M} (t) \rightarrow {\bf M} (t+\delta t)$ \cite{Palacios, Scholz, Nowak}. 
The Heun scheme is a good compromise between numerical stability 
and computational complexity.
In general, the statistical error of Heun scheme is made 
arbitrarily small by averaging over a sufficiently large number of 
stochastic trajectories. We do not carry out 
$\delta t \rightarrow 0$ limiting procedure but we employ 
a small discretization time interval, i.e., we use 
$\delta t=0.3 ps$ throughout the paper. 

\section{Results}

Many interesting features have already been shown by simply 
assuming the free layer is a single domain. As studied by 
Sun \cite{Sun2} and Bazaliy et al. \cite{Bazaliy2}, the
spin torque term in LLG makes the magnetization dynamics quite
complicated. The complication arises from the observation of the LLG
equation that the spin torque is fundamentally
different from the effective field term and from the damping term. 
The effective field term, first term in Eq.~(1), can be derived 
from the energy derivative with respect
to the local magnetization vector, i.e., ${\bf H}_{eff} =
- \frac{\partial E({\bf M}({\bf r}))}{\partial {\bf M}({\bf r})} $ where
$E({\bf M}({\bf r}))$ is the magnetic energy including the exchange energy
between the neighboring cells, the magnetostatic (dipole) energy, the 
(uniaxial) anisotropy energy, and the Zeeman energy. 
The damping term, which can not be
written as the energy derivative, selects the magnetization path such that
the local magnetization always moves into lower energy states, i.e., the
system is looking for an energy minimum. The spin torque neither behaves 
as an effective field which conserves magnetic energy nor the damping term
which dissipates the energy during the motion of magnetization.
The spin torque can increase or decrease the magnetic energy.
This leads to some interesting solutions with stable precessional states. 
We will show that these stable precessional states exist for finite
temperature and for non-single-domain structures.

Another interesting effect of the spin torque is the appearance of
the inverse hysteresis loop for certain range of the magnetic field,
i.e., the magnetization along the direction of the magnetic decreases
as one increases the magnetic field. To see this, 
let us consider the time-independent solution, $\frac{d\bf M}{dt} =0$.
Equation (1) becomes
\begin{equation} 
{\bf \Gamma} \equiv {\bf M} \times {\bf H}_{eff}- \frac{a_J}{M_s} 
{\bf M} \times ({\bf M} \times \hat{\bf M}_p) =0.
\end{equation}
For a single domain, we explicitly write 
${\bf H}_{eff}=(H_{ext}+H_{K} M_{x}){\bf e}_x
-4\pi M_{s} M_{z} {\bf e}_z$ and $\hat{\bf M}_{p}={\bf e}_x$. By placing
these expressions into Eq.~(2), we find
\begin{equation}
M_{z}[(H_{ext}+H_{K} M_{x})^{2}+4 \pi M_{s} M_{x}(H_{ext}+H_{K}
M_{x})+a_{J}^{2} M_{x}^{2}]=0
\end{equation}
The above equation indicates that $M_z =0$ is always a solution, i.e.,
the magnetization is in the plane of the layer. However, there are other
possible solutions with $M_{z} \neq 0$. From Eq.~(3), we have
\begin{equation}
(H_{ext}+H_{K} M_{x})^{2}+4 \pi M_{s} M_{x}(H_{ext}+H_{K} 
M_{x})+a_{J}^{2} M_{x}^{2}=0.
\end{equation}
Thus two solutions can be immediately identified,
\begin{equation}
H^{\pm}_{ext}=-H_{K} M_{x}-2 \pi M_{s} M_{x} \mp \sqrt{(2 \pi 
M_{s} M_{x})^{2}-a_{J}^{2} M_{x}^{2}}.
\end{equation}
For $2 \pi M_{s} \gg a_{J}$, two solutions are further simplified as 
$H^{+}_{ext}=-M_x (H_{K}+4 \pi M_{s} - \frac{a_{J}^{2}}
{4 \pi M_{s}}) $ and 
$H^{-}_{ext}= -M_x (H_{K}+\frac{a_{J}^{2}}{4 \pi M_{s}}) $.
These solutions show that the magnetic field and the magnetization
$M_x$ have opposite sign, i.e., an inverse hysteresis. 
However, we need to point out that these solutions do not always
stable compared to the simple solution $M_z=0$. When one examines the
stability condition, it is found that the above solutions 
$H^{\pm}_{ext}$ are possible for a finite range of $a_J$ \cite{Bazaliy2}. 

The third interest feature is that there is a critical current 
density, or a critical value of $a_J$, above which the spin torque 
can be used to
switch one magnetic configuration to another, for example, to
switch the magnetization from parallel to antiparallel alignment of the
two magnetic layers. If one assumes the single domain of the free layer,
the critical value can be readily deduced from the average energy 
variation rate \cite{Slonczewski,Sun2}, 
\begin{equation}
(a_J)_{crit} = \pm \alpha (2 \pi M_s + H_K ) + \alpha H_{ext}
\end{equation}
This simple relation indicates that
the current-magnetization hysteresis loop displays a jump when the 
current is swept through the critical current. 
One might raise a question that why the critical current
depends on demagnetization factor $4 \pi M_s$ in Eq.~(6).  
The answer is that the magnetization reversal involves  
a significant out-of-plane component of the magnetization. Once the
magnetization rotates out of the plane during a reversal process, the 
demagnetization factor is added to the effective field in LLG equation.
Thus, the damping term [$\alpha {\bf M} \times ({\bf M}\times {\bf H}_{eff})]$,
which is also proportional to the effective field, contains the demagnetization
factor. To overcome the damping, the critical spin torque is, therefore,
proportional to the demagnetization factor.  

The above features derived from the single domain structure 
show unique characteristics of the spin torque.
To further explore the spin torque effect, one needs to go beyond the
hysteresis analysis, i.e., one should look at the 
detailed magnetization dynamics,
and one must relax the assumption of the single domain. We now present  
our results in the following sections.

\subsection{Hysteresis loops with spin torques}

Without the spin torque, the hysteresis loop 
is almost a perfect square with the coercive field $H_c = H_K$, 
i.e., the hysteresis loop behaves as a Stoner-Wohlfarth particle
with the applied magnetic field parallel to the easy axis. 
When a spin torque is added on, the hysteresis loops display three 
distinct features. For a small current density, 
$|a_{J}| \ < 2\pi \alpha M_{s} $, the hysteresis loops are not 
affected by the spin torque, see Fig.~2a. When the current density
increases to intermediate values, the hysteresis loops begin to show 
some precessional states; those states are shaded in black, indicating
that the magnetization is never converged to a final
fixed direction; instead, it goes into a stable precession and we will
further discuss those precessional states in later sessions.
With this intermediate strength of the spin torque, the
precessional magnetization is oscillating around the easy axis. 
Just before the appearance
of the precessional states in Fig.~(2b), two irreversible jumps 
$H^{-}$ and $H^{+}$ occur; they can be again understood from the single 
domain solution. From the expression below Eq.~(5), one can identify 
\begin{equation}
H^{-}=-H_K-\frac{a_{J}^{2}} {4\pi M_{s}}
\end{equation}
On the other hand, for the field swept from negative $x$ to positive $x$ 
direction, one can use Eq.~(6) to solve for $H^{+}$ 
\begin{equation}
H^{+}=-\frac{|a_{J}|}{\alpha}+H_{K}+2\pi M_{s}
\end{equation}
We note that $H^{+}$ defined here is different from $H^{+}_{ext}$ of Eq.~(6)
because the latter is for a high magnetic field. In the present case, the
irreversible jump at $H^{+}$ occurs at much smaller field.
From Eqs.~(7) and (8), the coercivity $H_{c} = (H^{+}-H^{-})/2 
= H_{K}-\frac{1}{2}(\frac{|a_{J}|}{\alpha}
-2\pi M_{s})+\frac{a_{J}^{2}}{8\pi M_{s}}$ decreases with the spin torque and
the loops shift $\delta H = (H^{+}+H^{-})/2 = 
-\frac{1}{2}(\frac{|a_{J}|}{\alpha}-2\pi
M_{s})-\frac{a_{J}^{2}}{8\pi M_{s}}$ appears, i.e., 
the spin torque supplies a bias field to the loop. 

When the current density is further increased such that 
$a_{J} > \alpha (2 \pi M_{s}+2H_{K}$),
the loops become dramatically different from those of low current densities.
The loops show multiple jumps and
the precessional solutions expand to a large range of the magnetic field
(note the scale difference in Fig.~2c). In addition, there is the region
in the loop where the magnetization increases with decreasing magnetic field,
e.g., when the magnetic field decreases from -9,900 (Oe) to -12,4000 (Oe),
the magnetization $M_{x}$ increases from 0.78 to 0.99. This 
unusual solution at high field is consistent with the single domain solution
given by Eq.~(4), i.e., we can solve for $M_{x}$ from Eq.~(4), 
\[
M_{x}=\frac{-H_{ext}(H_K+2\pi M_s - \sqrt{H_{K}^{2}
+(2\pi M_{s})^{2}-a_{J}^{2}}}{H_{K}^{2}+4 \pi M_{s} H_{K}+a_{J}^{2}}. 
\]
The solutions are stable; we have varied temperatures and 
confirmed that these states exist for all temperatures.

\subsection{Magnetization-current hysteresis loops at finite external fields}

In Fig.~3, we show M-I hysteresis loops (magnetization as a function of
the spin torque) for several different external fields. 
When the external magnetic field is smaller than the anisotropy
field, i.e., $H_{ext} < H_{K}$, the M-I loop is almost square in shape
and the critical spin torque is very close to the
analytical expression given by Eq.~(6). For example, the loop shift is
$\delta a_J=-6$ (Oe) for the external field of 200 (Oe), 
see Fig.~3a. 
When the external field is in the range between $H_{K}$ and $4\pi M_{s}$,
the precessional solutions begin to appear. Further increasing the
(negative) spin torque results a new stable state, see the plateau
labeled B in Fig.~3b. This new stable ``B'' state exists even for
a single domain case as pointed out in Ref. \cite{Bazaliy2}. 
If one applies an external field larger than $4\pi M_s$, the precessional
states disappear for the M-I loops.
Instead, there are multiple minor loops at
large current density, see Fig.~3c.

\subsection{Energy dissipation and energy pumping}

The new stable states, labeled as ``B'' in Fig.~3b, deserve further
investigation. This is because the existence of the ``B'' state is yet another 
signature of the effect due to spin torque; without the spin torque, the 
magnetization of the
final stable state would be in the plane for any in-plane magnetic field
and the torque ${\bf \Gamma}={\bf M} \times {\bf H}_{eff}$ equals to zero.
In Fig.~4, we show the magnetization vector pattern for the ``B'' state: 
there is a significant out-of-plane component. If we calculate the
effective magnetic field ${\bf H}_{eff}$ for the ``B'' states, we  
find that the effective field is {\em not} parallel to the
magnetization direction, there are angles between the two vectors 
${\bf H}_{eff}$ and ${\bf M}$ for those ``B'' states. 
This would be unacceptable for the LLG without the spin torque, 
because the non-parallel 
configuration between ${\bf H}_{eff}$ and ${\bf M}$ can 
not be in the static stable condition. 
However, in the presence of spin torque it is possible, because the 
static stable condition is 
zero total torque, ${\bf \Gamma}_{tol}={\bf M}
\times {\bf H}_{eff}- \frac{a_J}{M_s} {\bf M}\times ({\bf M}
\times \hat{\bf M}_p )=0 $. 

Another interesting observation is the variation of the magnetic energy
of the system. Without the spin torque, the  
change of the magnetic energy of system can be easily derived from LLG
equation, 
\begin{equation}
\frac{d E}{d t} = -\frac{\alpha \gamma}{1+\alpha ^2} \frac{1}{M_s}
| {\bf H}_{eff} 
\times {\bf M} | ^{2}.
\end{equation}
Thus, the non-parallel configuration between ${\bf H}_{eff}$ and ${\bf M}$
would continuously dissipate the energy, i.e. $\frac{d E}{d t}<0$, 
until the magnetic energy of system reaches the local minimum, i.e.,
${\bf H}_{eff} //{\bf M}$.
When the spin torque turns on, it is possible to 
compensate the loss of the energy by the gain from 
the spin torque,
\begin{equation}
\frac{d E}{d t} = -\frac{\gamma}{1+\alpha ^2}
\frac{1}{M_s}
\left[ 
\alpha | {\bf H}_{eff} \times {\bf M} | ^{2} -
a_{J} (\alpha M_s \hat{\bf M}_{p}- {\bf M} \times \hat{\bf M}_p) \cdot
( {\bf H}_{eff} \times {\bf M} )
\right]
\end{equation}
where the first term is the energy loss due to damping and the second
term is the energy input (output) due to spin torque. If the first term is 
less than the second, i.e. $\frac{d E}{d t}>0$, the magnetic energy increases. 
If the first term 
is exactly balanced with the second, the net energy loss will be null and
it is therefore possible to form a stable state ``B'' where 
$|{\bf M} \times {\bf H}_{eff}|$ is nonzero but $\frac{d {\bf M}}{dt}=0$.
If $\frac{d E}{d t}$ becomes a periodic function,
a stable precessional state appears. 

\subsection{Precessional states}

As we have shown in the preceding sections, the solution of the LLG
contains stable precessional states in which the magnetization never
converges to a final state as if the system had no damping at all.
These precessional solutions are one of the unique properties
of the magnetization dynamics driven by a spin-polarized current.
Without the current, the system is always losing its energy
due to Gilbert damping and thus the precession can not be 
sustained after certain time scales, typically in a few nanoseconds. 
In Fig.~5, we show two trajectories of magnetization vectors 
at the stable precessional states for two magnetic fields but the same current.
These two trajectories clearly show a large out-of-plane component 
of magnetization, indicating the significant role played by the demagnetization
factor. 

A further inspection of the magnetic energy of the precessional states
reveals interesting temporal variation of the energy. The system constantly 
gains energy via the
transfer of the angular momentum of the conduction electrons (spin currents)
to the local magnetization. Thus the precessional states are the result of 
competition between the Gilbert energy dissipation and the spin torque energy 
input. The damping and pumping rates are not the 
same at a given time. The magnetic energy 
oscillates even for a steady state current. In Fig.~6, 
we show the evolution of magnetic energy 
as a function of time for different damping
constants. The initial state of the system is a ``leaf state'' 
without the external field and the spin torque. At t=0, we simultaneously
apply an external field $H_{ext} = -2000$(Oe) and $a_J=-400$ (Oe). 
For a large damping constant, i.e.,  $\alpha > \frac{|a_{J}|}
{2\pi M_{s}}$, the energy dissipation is dominant. Thus, the system
loses magnetic energy and eventually sets into a local minimum energy
as shown in Fig.~6a. With decreasing $\alpha$, for example $\alpha=0.03$,
the energy dissipation by damping and energy pumping by the spin torque
becomes comparable. In response to the applied field and the spin torque, the
magnetic energy begins to oscillate in a precessional cycle. If the damping
constant further decreases, the energy oscillates around a value that
increases and eventually reaches an asymptotic value. Such a dynamic
change of the magnetic energy can be explained again from Eq.~(10).
When ${\bf M}_{p}$
is parallel to ${\bf H}_{eff}$, the rate of the energy change 
reduces to pure dumping or negative damping, i.e.,
\begin{equation}
\frac{d E}{d t}= - \frac{\gamma}{1+\alpha ^2} \frac{1}{M_s} \left(
\alpha - \frac{a_J}{|H_{eff}|}
\right)  
| {\bf H}_{eff} \times {\bf M} |^{2}. 
\end{equation}
If $\alpha - \frac{a_J}{|H_{eff}|} >0$ the magnetization seeks for 
energy minimum, otherwise, it seeks for the energy maximum, i.e., 
negative damping. In general, however, ${\bf M}_{p}$
is not parallel to ${\bf H}_{eff}$ and thus, Eq.~(10) can be positive and
negative at different times and one arrives at the precessional motion.
 
It is interesting to observe that the frequency of the stable precession
highly depends on the damping parameter; this is in sharp contrast with
the normal precession where the frequency is determined by the effective
field. The reason is rather simple: the stable precessional states
described here are from the competition between the damping and spin torque,
and they are not the initial precessional motion (which is unstable) 
driven by the effective field. 
In Fig.~7, we show the precessional frequencies as a function of damping 
constant $\alpha$. As expected, the frequency monotonically
increases as one decreases the damping parameter. For the same 
reason, the frequency increases with increasing spin torques. 

To further analyze these precessional states, we have shown, 
in Fig.~8 the temporal evolution of three magnetization components when the 
magnetization
reaches a stable precessional state. The magnetization shows significant
out-of-plane component (z-direction) during the oscillation. The swing 
of magnetization in the in-plane hard axis direction is as large as that in the
easy axis direction. A noticeable difference between low temperature and room
temperature is the difference for the out-of-plane component of magnetization.
There are two stable precessional solutions at the given external
field and spin torque. One is the precession at the upper half plane, i.e.,
$M_z > 0$, as shown in Fig.~8a. Another solution is located at
the lower half plane, i.e., we simply replace $M_z$ by $-M_z$ and keep the
$x$ and $y$ components as in Fig.~8a. These are degenerated solutions
and the magnetization dynamics takes either one of the two solutions but
not both at zero temperature.
At room temperature, however, these two solutions can jump around, making
the out-of-plane component crossing the x-y plane. This thermally activated
transition between two precessional states originates from low barrier
heights--the thermal energy at room temperature can not be ignored
in this small structure, as shown in Fig.~8b.   
While the thermal energy can not wash out the precessional solutions,
it does alter the magnetization dynamics significantly.

\subsection{Switching speed}

We have shown that there is a critical spin torque to switch
the magnetization. We now address how fast the switching is. 
The switching speed has been analytically calculated for a single
domain sphere and a single domain thin film, in the absence of the
spin torque, i.e., switching by the external field. 
The speed depends on the damping parameter. The optimal damping 
parameter for the fastest switching 
is $\alpha =1$ for the sphere and $\alpha \approx 0.013$ for the 
thin film \cite{Kikuchi}.
Our question is what is the optimal damping parameter for the spin torque
induced switching? How fast the switching speed compared with the 
field induced switching?

Let us first take a look at the influence of the spin torque on the
switching speed by a magnetic field. The initial magnetization 
(``leaf state'') is at the
positive x-direction. A reversed magnetic field is applied in the negative
x-direction at $t=0$. When the x-component of the 
magnetization $<M_{x}>$ reaches 
at least -0.95 and it stays below this value after $t_s$, we define this
$t_s$ as the switching time. In Fig.~9, we show the dependence of the
switching time for three different spin torques. The switching time is
enhanced for one direction of the current and is shortened for the
opposite direction. We have noticed that the optimal damping parameter 
varies when the spin torque changes. 

To study the switching speed due to spin torque, we 
must apply a spin torque exceeding the critical spin torque. 
In Fig.~10, we show the switching time 
after a spin torque larger than the critical spin torque is applied. The
switching time is very slow as the spin torque approaches the
critical spin torque. We find that the switching time can be reasonably 
fitted by $t_s^{-1} \propto = a_J -a_{crit}$. Since the critical spin torque
$a_{crit}$ is proportional to the damping parameter, the applied spin
torque is shifted for different damping parameters as shown in Fig.~10.

Up till now, we have concentrated our description of the spin torque
in the limit of the steady state current. It is, however, desirable to
see the magnetization dynamics in the case of a pulsed form. Here we
should consider a simple form of the spin torque pulse: a square pulse whose 
width is $t_w$, i.e., we will ignore the
rise and fall times of the pulse; in real devices, one should also consider 
these times. For such a pulsed electrical current, one would expect
that whether the magnetization can be switched by the spin torque depends
on pulse's amplitude {\em and} width. For a fixed pulse amplitude
larger than the critical spin torque, there is minimum pulse width  
needed to achieve current-driven magnetization reversal. 
In Fig.~11, we show the magnetization dynamic after the pulse
of $a_J =2500$ (Oe) is applied at $t=0$. 
In the first two panels, the durations of the pulse are just not 
enough to fully reverse the magnetization,
i.e., the magnetization returns to the initial state ($<M_{x}> \approx 1.0$)
after one removes the current. 
Just a small increase of the pulse width, see Fig.~11c, the magnetization
is able to fully reverse itself after the signal is taken away. 
Thus, $t_w = 0.775$ (ns)
is defined as the minimum width required for the current-induced 
magnetization reversal for the pulse amplitude $a_J =2500$ (Oe). 

The minimum pulse width defined above depends on the pulse amplitude.  
With increasing pulse amplitude, the minimum pulse width 
decreases. We can construct a phase diagram of magnetization
reversal by varying the pulse amplitude and repeating the procedure to
determine the minimum pulse width in each case. As shown in Fig.~12, 
the minimum pulse width increases dramatically for the 
pulse amplitude close to the critical spin torque; this phenomenon is
very similar to the classical ``critical slowing down'' in statistical
physics: when an external force approaches a critical value, the system
slows down. In our case, we find that the 
reversal time is mainly spent at the beginning stage of switching. 

The minimum pulse width is also affected by other parameters. In Fig.~13,
the computed reversal/non-reversal phase diagram is plotted in terms of 
the damping constant for several external magnetic fields. 
Since the thin film anisotropy field is about $500$ Oe, 
the largest external fields ($|H_{ext}| \leq 200 $ Oe) used in Fig.~13
can not trigger the magnetization reversal itself. 
The increase of the minimum width for a larger damping constant is consistent
with the picture that the spin torque is competing with the damping process.
One interesting feature in Fig~13 is that the external field affects
the minimum width much more significantly for large damping parameters
than for small ones. This is due to the fact that the spin torque
($a_{J}=2500$ Oe) is closer to the critical spin torque
for larger damping constant. By applying the magnetic field, one can
shift the critical spin torque to smaller values. Therefore, the minimum width
is much smaller compared to that without the magnetic field. 

\subsection{Thermal switching}

Finally, we address the problem of thermally activated
magnetization switching in the
presence of the spin torque. As the device approaches nanometer
size, temperature driven magnetization reversal becomes one of
key factors limiting the device performance\cite{Koch, Rizzo, Wernsdorfer}. 
We have already seen in Fig.~8b 
that the inclusion of the thermal activation has significantly 
altered the magnetization dynamics. 

The thermally assisted magnetization reversal is conventionally modeled by
the N\'{e}el-Brown formula $\tau = f_0^{-1} \exp (E_b/k_BT)$ where 
$\tau$ is the thermal switching time, $E_b$ is the energy barrier, 
and $f_0$ is the attempt frequency of the order of $10^9$ (s$^{-1}$). 
For a single domain particle, the energy
barrier is simply $E_b =E_0(1-H/H_c)^{\beta}$ where $H_c$ is the coercive
field and $E_0$ is an extrapolation of the energy barrier at zero field. 
Notice that the energy
barrier is independent of the damping parameter. If the spin torque is purely
competing with the damping, one would expect that the spin torque does not
affect the energy barrier and thus the component $\beta$ would be
independent of the spin torque. We have found, however,
the spin torque does more than altering the damping parameter as shown below.    
Let us suppose the magnetization of the free layer is initially 
saturated in the positive $x$ direction. At $t=0$, we 
apply a negative magnetic field which is close to but less than the switching 
field $H_{c}$. At the same time, a spin torque is also applied to the system.  
At the finite temperature, the average waiting time for the magnetization 
reversal is given by the Arrhenius law, the switching probability decays as 
$e^{-t/\tau}$. To minimize the statistic error,
we determine the relaxation time (average waiting time)
by repeating the above procedure 
$800$ times for each spin torque. In Fig.~14, we show  
the probability of the free layer not being switched ($1-P_{s}^{exp}(t)$) as
a function of the relaxation time $t$ for a magnetic field $H_{ext}=-438$ (Oe). 
The reversal probability $P_{s}^{exp}(t)$ increases with 
spin torque; the positive spin torque ($a_J > 0$) leads to a faster 
thermal switching and the negative spin torque results in slower 
thermal reversal. The distribution of the relaxation time can be well-fitted 
to the Arrhenius formula function, but now the energy barrier depends on the
spin torque: the positive current favors lower energy barrier.
We show the fitted relaxation time as a function of the spin torque
for two different external fields in Fig.~15. The linear relation indicates 
the effective energy barrier is linearly dependent on the current.
Such results can be potentially very useful: one can control the energy
barrier of a nanoscale magnetic element by applying a proper spin polarized 
current so that superparamagnetic effects can be overcome.

\section{Discussions and Conclusions}
The spin torque in LLG equation adds a new degree of freedom to control
and manipulate magnetization dynamics. The spin torque differs from
the other common torques: the one due to the effective field and the
another one due to the dissipation (Gilbert damping). The torque from the
effective field is a conservative torque in the sense that the magnetic
energy is conserved while the dissipative torque produces energy dissipation
as long as the magnetization is not static. The spin torque can play the role
of the effective field as well as the dissipation torques; it is this peculiar
feature making the dynamics and hysteresis loops quite unusual. We now
summarize our results below.

First, in our thin sample with the dimension of
64nm by 64m by 2.5nm, the hysteresis loops
and the dynamics behave qualitatively as a single domain. For 
example, the magnetization of each grid never differs by more than $5^o$
during any dynamic processes of the magnetization reversal (except at very
high magnetic fields, $H_{ext}> 4 \pi M_{s}$ when the spin wave excitations
are generated; we will address the effect of extremely high field elsewhere). 
Therefore, most
of the conclusions obtained from the analytical work where the layer is
treated as a single domain are justified in the present study. However,
the non-single domain feature is important when one addresses the questions
such as the reversal speed. For example, 
in the single domain picture, one 
would never switch the magnetization by a large magnetic field or a large
spin torque if the field and the magnetization directions are exactly at the
easy axis, because there will be no initial torque in this case. If one
takes the full micromagnetics into account, the magnetization 
at the center of the sample
differs slightly from that at the edges with some characteristic
frequencies. Such slight non-uniformity of the magnetization is just
enough to allow spin-torques to reverse magnetization. 
Therefore, one should be cautious in utilizing the single
domain picture in addressing the detailed dynamics of magnetization reversals.

The second conclusion we wish to emphasize is the unique feature of the
spin torque: the new equilibrium states and the stable precessional states. 
These states would not be
stable without the spin torque since the energy would be damped as long as
the magnetization vector is not parallel to the direction of
the effective magnetic field. This is an analogy to the classical moving
object: the object will eventually come to rest if no power is added 
to the object because air resistance would consume all the kinetic energy of
the object. Here, the spin torque is an external power supplied to the
system. With appropriate
conditions, i.e, with proper magnetic fields and damping constants,
the stable precessional states appear, which is again analogous to 
the ``resonant'' states of classical mechanics. We have found that these
stable precessional states are indeed very stable: they exist even if there
is large thermal fluctuation and they appear to exist for a wide range of
parameter space. 

Next we have studies the influence of the spin torque on switching
speed. We have found that the spin torque can enhance and retard the
switching speed. The enhancement is large, usually a few times faster.
If one wishes to have an order of magnitude increase, one should consider
the spin torque and magnetic field applied in other directions, not
the direction of the easy axis. It has been known that the hard axis 
application of these components can dramatically increase the switching
speed. One can easily explore the best scenario for the ultra fast
switching devices by changing directions of the spin torque. 
Finally, we have pointed out that the effective energy barrier for thermally
assisted reversal can be influenced by the spin torque. This result further
illustrates the fundamental difference between the spin torque and
the damping. 

In conclusion, we have shown the various dynamic behaviors in the
presence of the spin torque. We hope that this work can generate 
more experimental efforts to verify those novel dynamic phenomena 
unique to the spin torque. This work is supported by NSF (ECS-0223568)
and INSIC.

\pagebreak

\begin{figure}
\caption{Schematic of the pillar device with a positive spin current 
defined as the electrons flowing from the thinner to the thicker 
$Co$ layers (opposite direction for the current flow).}
\end{figure}
\begin{figure}
\caption{Hysteresis loops of the free $Co$ layer in the 
presence of the different spin torques: (a) $a_J=-100$ (Oe); 
(b) $a_J =-200$ (Oe); (c) $a_J=-400$ (Oe).}
\end{figure}
\begin{figure}
\caption{M-I loops with different external fields: (a) $H_{ext}=-200$ Oe;
(b) $H_{ext}=-2000$ Oe; (c) $H_{ext}=-20000$ Oe.}
\end{figure}
\begin{figure}
\caption{Magnetic moments at the point ``B'' in Fig.3(b), where 
$H_{ext}=-2000 $ Oe, $a_J =-3000$ (Oe).}
\end {figure}
\begin{figure}
\caption{The trajectories of two stable precessional states
that are within the shaded area of Fig.~2c, where $a_{J}=-400 $ Oe.}
\end {figure}
\begin{figure}
\caption{Temporal evolution of the magnetic energy 
for three different damping constants $\alpha$ ($=0.1, 0.03, 0.008$). 
The external field is $H_{ext}=-2000 $ Oe and spin torques 
$a_J=-400$ (Oe).}
\end{figure}
\begin{figure}
\caption{Precessional frequencies $\omega$ as a function of damping  
constant $\alpha$ for different $H_{ext}$ and $a_J$: $H_{ext}=-1000$ Oe,
$a_J=-500$ (Oe) (open triangle); $H_{ext}=-2000 $ Oe,
$a_J=-500$ (Oe) (open circle); $H_{ext}=-1000 $ Oe,
$a_J=-1000$ (Oe) (solid circle).}
\end{figure}
\begin{figure}
\caption{Temporal evolution of the total magnetization components for a 
stable precessional state. (a) T=0 K; (b) T=300 K. The magnetic field is 
taken as $-2000 $ Oe and spin torque $a_J=-400$ (Oe)
that is within the shaded area of Fig.2(c).}
\end{figure}
\begin{figure}
\caption{Swithing speed vs damping constant $\alpha$ for three spin 
torques $a_{J}=+100$ (Oe) (dashed line), $a_J =0$ (solid line) and 
$a_J =-100$ (Oe) (dotted line).
The external field is taken as $-510 $ Oe.}
\end{figure}
\begin{figure}
\caption{Switching speed vs spin torque (larger than the 
critical spin torque) at $H_{ext}=0.0$ (Oe) for different 
damping constants.}
\end{figure}
\begin{figure}
\caption{Time evolution of the magnetization component $M_{x}$ with 
three different pulsed torque widths (0.75 ns, 0.765 ns and 0.775 ns).
The pulsed amplitude is $2500$ (Oe) and 
$H_{ext}=0.0 $ (Oe).}
\end{figure}
\begin{figure}
\caption{Minimum pulsed spin torque width versus amplitude 
for zero external field.}
\end{figure}
\begin{figure}
\caption{Reversed/not-reversed boundaries as a function of 
damping parameter for different external 
fields: $H_{ext}=0$ Oe (open triangle), 
$H_{ext}=-100$ Oe (full square) and $H_{ext}=-200$ Oe (open circle). 
The pulsed amplitude is taken as 2500 (Oe).}
\end{figure}
\begin{figure}
\caption{Probability of not being switched magnetization as a function
of time at T=300 K for different spin torques.
The solid lines are fits to the function $e^{-t/ \tau}$.}
\end{figure}
\begin{figure}
\caption{The relaxation time $\tau$ (solid line) 
and the switching probability at $t=50ns$ (dashed line) as a function 
spin torque at T=300 K for two different external fields.}
\end{figure}

\end{document}